\documentclass[namedreferences]{solarphysics}

\usepackage[hyperref,optionalrh]{spr-sola-addons} 
\usepackage{graphicx}        
\usepackage{color}           

\usepackage[utf8]{inputenc}
\usepackage{enumerate}
\usepackage[makeroom]{cancel}

\chardef\us=`\_

\begin{document}

\begin{article}
\begin{opening}

\title{How Magnetic Erosion Affects the Drag-Based Kinematics of Fast Coronal Mass Ejections}

\author[addressref={aff1},corref,email={s.stamkos@uoi.gr}]{\inits{S.}\fnm{Sotiris}~\lnm{Stamkos}\orcid{0000-0003-2798-2801}}
\author[addressref=aff1,email={spatsour@uoi.gr}]{\inits{S.}\fnm{Spiros}~\lnm{Patsourakos}\orcid{0000-0003-3345-9697}}
\author[addressref=aff2,,email={Angelos.Vourlidas@jhuapl.edu}]{\inits{A.}\fnm{Angelos}~\lnm{Vourlidas}\orcid{0000-0002-8164-5948}}
\author[addressref=aff3,,email={iadaglis@phys.uoa.gr}]{\inits{I.A.}\fnm{Ioannis A.}~\lnm{Daglis}\orcid{0000-0002-0764-3442}}

\address[id=aff1]{Department of Physics, University of Ioannina, Greece}
\address[id=aff2]{Johns Hopkins University, Applied Physics Laboratory, Maryland, United States}
\address[id=aff3]{Department of Physics, National and Kapodistrian University of Athens, Greece}

\runningauthor{S. Stamkos et al.}
\runningtitle{Magnetic Erosion and CME Kinematics}

\begin{abstract}
In order to advance our understanding of the dynamic interactions between coronal mass ejections (CMEs) and the magnetized solar wind, we investigate the impact of magnetic erosion on the well-known aerodynamic drag force acting on CMEs traveling faster than the ambient solar wind. In particular, we start by generating empirical relationships for the basic physical parameters of CMEs that conserve their mass and magnetic flux. Furthermore, we examine the impact of the virtual mass on the equation of motion by studying a variable-mass system. We next implement magnetic reconnection into CME propagation, which erodes part of the CME magnetic flux and outer-shell mass, on the drag acting on CMEs, and we determine its impact on their time and speed of arrival at 1 AU. Depending on the strength of the magnetic erosion, the leading edge of the magnetic structure can reach near-Earth space up to $\approx$ three hours later, compared to the  non-eroded case. Therefore, magnetic erosion may have a significant impact on the propagation of fast CMEs and on predictions of their arrivals at 1 AU. Finally, the modeling indicates that eroded CMEs may experience a significant mass decrease. Since such a decrease is not observed in the corona, the initiation distance of erosion may lie beyond the field-of-view of coronagraphs (i.e. 30 $\mathrm{R_{\odot}}$).
\end{abstract}
\end{opening}


\section{Introduction} 

A coronal mass ejection (CME) is the release of a significant amount of magnetized plasma from the solar corona that moves away from the Sun, and once it reaches the interplanetary medium can then be measured as an interplanetary CME (ICME). Although the magnetic field plays the most important role in the initiation and early evolution of CMEs, there are many studies suggesting that approximately from $15~\mathrm{R_{\odot}}$ outwards, the Lorentz forces are negligible and it is the momentum coupling between a CME and the solar wind via a drag force that dominates their dynamics   
\citep[e.g.][]{2001JGR...10629207G, 2006SoPh..233..233T, 2015ApJ...809..158S}.

Drag models are based on the concept of magnetohydrodynamic (MHD) drag, which, in contrast to the kinetic-drag effect in a fluid, is supposed to be caused primarily by the emission of MHD waves in the collisionless solar-wind environment \citep{1996JGR...101.4855C}. Observations of slow (fast) CMEs acceleration (deceleration) towards the ambient solar-wind speed, led to the conclusion that it is the drag force that is responsible for the relative equalization of  CME and solar wind speeds \citep[see][]{2010A&A...512A..43V, 2010ApJ...717L.159P} and therefore drag-based models could be used for the prediction of the arrival times of CMEs at various locations in the inner heliosphere and beyond. \cite{2007A&A...472..937V} proposed that the equation describing the aerodynamic drag can be utilized to establish a simple drag-based model of CME propagation. Although the drag force between the CME and the solar wind is rather  well established and can be quite easily modeled, several effects such as CME deformation, front flattening, deflection, rotation, erosion, and expansion can be relevant for CME propagation  \citep[see, for example, the review by][]{2017SSRv..212.1159M}.

Understanding the interactions between CMEs and the ambient solar wind has presented the scientific community with formidable challenges. Predicting the time and speed of arrival of CMEs (ToA and SoA, respectively) in the near-Earth space environment, or other locations in the inner heliosphere, using apart from drag-based modeling (DBM) \cite[e.g.][]{2014ApJS..213...21V, 2015ApJ...806..271S, 2018ApJ...854..180D}, a variety of other methods (e.g. empirical, physics-based, time-dependent MHD), frequently leads to significant errors  \citep[e.g.][]{2019RSPTA.37780096V}. The complexity of CME\,--\,solar wind interactions, lack of critical observations of CMEs and the solar wind, and gaps in theory/modeling are examples of factors that induce significant difficulties in predictive schemes of CME impacts. With all these methods considered, the mean absolute error (MAE) for predicted ToAs of CMEs at 1 AU is greater than 12 hours \citep[e.g.][]{2019RSPTA.37780096V}.  

\begin{figure}[h]
\centering
\includegraphics[width=12cm]{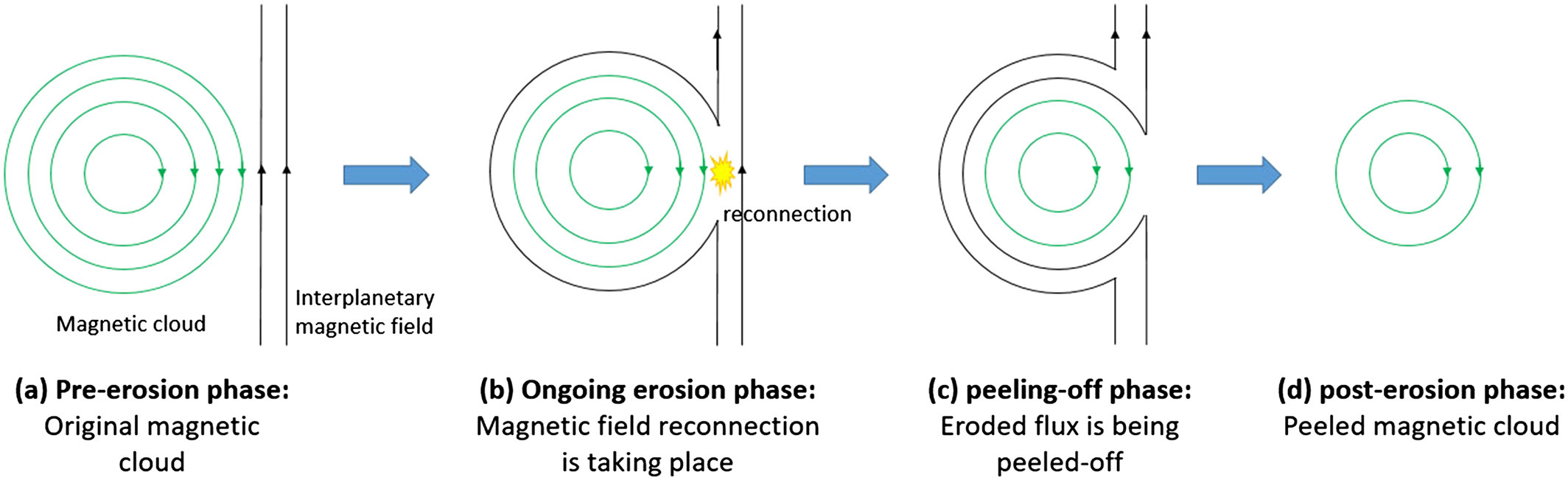}
\caption{Magnetic erosion applying to a CME. Magnetic erosion is  due to magnetic reconnection between oppositely directed CME and IMF magnetic fields for a cylindrical CME (Panels \textbf{a} and \textbf{b}) and it extracts  concentric magnetic cells and the frozen-in mass from the outer shell of the CME (Panels \textbf{c} and \textbf{d}). Used with
permission of John Wiley \& Sons, from J. Geophys. Res., Understanding the twist distribution inside magnetic flux ropes by anatomizing an interplanetary magnetic cloud, Wang et al., 123, 3238, 2018; permission conveyed through Copyright Clearance Center, Inc.} 
\label{fig:1}
\end{figure}

Magnetic reconnection is ubiquitous in space \citep[e.g.][]{2007mare.book.....P}. It is important for the formation of most of the solar-dynamic phenomena and it has been detected from the low solar atmosphere up to the Earth’s magnetotail. Recently, magnetic reconnection has been shown to occur regularly in the solar wind \citep[e.g.][]{2005JGRA..110.1107G, 2013ApJ...763L..39G}. During its propagation in the interplanetary medium, a CME interacts with the interplanetary magnetic field (IMF) and magnetic reconnection may occur (see for example the schematic of Figure \ref{fig:1}). If it happens at the front boundary of a magnetic cloud (MC), i.e. a coherent CME structure resembling a magnetic-flux rope, characterized by enhanced magnetic-field intensity, a smooth rotation of its magnetic-field components, and a much lower proton temperature with respect to the background solar wind \citep{1982GeoRL...9.1317B}, then magnetic reconnection erodes part of its entrained magnetic flux and peels off the flux rope's outer layers \citep{2006A&A...455..349D}. Magnetic erosion leads to an imbalance of the azimuthal magnetic flux in the front and rear part of CMEs. This is indeed used as an observational signature of magnetic erosion \citep{2006A&A...455..349D}. 

\cite{2012JGRA..117.9101R} and \cite{2014JGRA..119...26L} examined several CMEs and found evidence that magnetic reconnection took place in their fronts and gave rise to magnetic erosion. In addition, \cite{2012JGRA..117.9101R} reported magnetic-flux decreases of $44\,\%$ and $49\,\%$ as the studied CME propagated to the \textit{Advanced Composition Explorer} (ACE) and the \textit{still-operational Solar Terrestrial Relations Observatory} (STEREO-A), respectively. \cite{2015JGRA..120...43R} performed a statistical study of 109 magnetic clouds observed by \textit{Wind}, 78 by STEREO-A, and 76 by STEREO-B during the period 1995–2012. Due to the importance of reliable boundary determination in the implementation of the analysis methods, they investigated in detail each event to define the magnetic-cloud boundaries with the best accuracy possible. They suggested that magnetic clouds may be eroded at the front or rear and in similar proportions, with significant average erosion of about 40\,\% of the total azimuthal magnetic flux. On average, 42\,\% of the magnetic clouds were eroded at the front and 33\,\% at the rear (relative to the total azimuthal flux content). Their results are consistent with the frequent (up to $\approx$ 30\,\%) observation of reconnection signatures locally at both the front and rear boundaries \citep{2010ApJ...720..454T}.

In this work, we incorporate magnetic reconnection caused by interactions between CMEs and the IMF \,--\, solar-wind couple, resulting in magnetic erosion, into the drag-based CME propagation model and examine its impact on the Time and Speed of Arrival of a CME structure at 1 AU. Section 2 contains our theoretical framework. In Section 2.1 we discuss the drag-based propagation of non-eroded CMEs and then in Section 2.2 we derive empirical profiles of the radial evolution of key physical parameters of non-eroded CMEs. In Section 2.3 we incorporate magnetic erosion into a new drag-based model of CME kinematics. In Section 3, we discuss our model's results for both eroded and non-eroded CMEs. Finally, in Section 4, we provide a summary of our work and discuss future plans. 

\section{Theoretical Framework}
This section contains our theoretical framework for the formulation of a drag-based model incorporating magnetic erosion.

\subsection{Aerodynamic Drag Force Acting on Non-Eroded CMEs} 

We start with a discussion of the drag force acting on non-eroded CMEs. Consider a cylindrical flux-rope CME with radius $R$ and height $L$ which radially propagates in the interplanetary medium. The assumption of radial CME propagation essentially leads to 1D equations for the structure's kinematics. The drag force [$F_{\mathrm{D}}$] acting on it, and in particular the associated rate of change of the velocity, is given by \cite{2004SoPh..221..135C}:

\begin{equation}\label{eq:1}
   \frac{\mathrm{d}V_{\mathrm{i}}}{\mathrm{d}t}=\frac{F_{\mathrm{D}}}{M_{\mathrm{tot}}}=-\gamma C_{\mathrm{D}} |V_{\mathrm{i}}-V_{\mathrm{e}}|( V_{\mathrm{i}}-V_{\mathrm{e}})
   \end{equation}

Subscripts i and e are used to represent quantities inside and external to the ICME with $V_{\mathrm{i}}$ and $V_{\mathrm{e}}$ corresponding to the bulk speed of the CME (i.e. the speed of the center of the cylindrical structure) and the ambient solar wind speed, respectively. $M_{\mathrm{tot}}$ is the total mass of the CME\,--\,solar-wind system and will be discussed in detail later in this section. The $\gamma$-parameter is an inverse deceleration length given by:

\begin{equation}\label{eq:2}
   \gamma=\frac{\rho_{\mathrm{e}}A}{\tau(\rho_{\mathrm{i}}+\rho_{\mathrm{e}}/2)}=\frac{4}{\pi R(2\rho_{\mathrm{i}}/\rho_{\mathrm{e}}+1)},
\end{equation}

where $\mathrm{\tau}$ and $\mathrm{A}$ are the volume and reference area (i.e. lateral cross-sectional area) of a cylindrical CME structure, respectively, and  $\rho_{\mathrm{i}}$ and $\rho_{\mathrm{e}}$ are the mass density of the CME and of the ambient solar wind mass density, respectively.

Finally, $C_{\mathrm{D}}$ is the drag-force coefficient, a dimensionless number that encapsulates all the complex dependencies of a given structure's drag on shape, inclination, flow conditions, etc. For the calculation of $C_{\mathrm{D}}$ we followed the formulation of \cite{2015ApJ...809..158S}, which is based on a microphysical prescription for viscosity in collisionless plasmas of the turbulent solar wind \citep{2012GeoRL..3919107S}. In our calculations of $C_{\mathrm{D}}$ we used the radial profile of proton density from \cite{2013JGRA..118.1351H}.

We now discuss the important aspects of the total mass [$M_{\mathrm{tot}}$] of the CME\,--\,solar wind system. We have that:

\begin{equation}\label{eq:3}
   M_{\mathrm{tot}}=M_{\mathrm{i}}+m_{\mathrm{virtual}},
\end{equation}

with $M_{\mathrm{i}}$ corresponding to the mass of the cylindrical CME. $m_{\mathrm{virtual}}$ is the so-called added mass or virtual mass of fluid dynamics \citep[e.g.][]{white2011fluid} applying when a body moves through a fluid. It relates to the inertia added to a system because an accelerating or decelerating body must displace some volume of surrounding fluid as it moves through it. In other words, virtual mass naturally arises, because the object and surrounding fluid cannot simultaneously occupy the same volume. The inclusion of the virtual mass leads to the requirement of an extra force to accelerate a body moving in a fluid compared to the case when it moves in a vacuum. This force is sometimes referred to as apparent mass force \citep{book:1197664} because it is equivalent to adding a mass to the existing body. 

The concept of virtual mass can be incorporated in the study of CME propagation \citep[e.g.][]{2004SoPh..221..135C, 2021JSWSC..11...34V}, where the moving CME body gives rise to accumulation or in other words to a pile-up of solar-wind mass around it. The piled-up, i.e. compressed plasma, is contained within the sheath formed around CMEs. 
 
Frequently, in applications of the drag-force model, the virtual mass is  neglected \citep[e.g.][]{2013SoPh..285..295V}. As we can see from Equation \ref{eq:2} when $\rho_{\mathrm{CME}} \gg \rho_{\mathrm{sw}}$ the virtual mass becomes negligible. \cite{2021JGRA..12628380T} combined remote-sensing and in-situ observations with the  Graduated Cylindrical Shell model \citep[see][]{2011ApJS..194...33T} and ascertained that the sheath region should be treated as a significant extra mass. An indication that the sheath becomes much more prominent in the interplanetary medium is also given by a relative increase of the sheath duration from Mercury to Earth \citep{2019JGRA..124..812J}.  

Obviously, the virtual mass corresponding to a CME increases with distance since more and more mass is piled up as it outwards propagates. We furthermore assume that the CME mass [$M_{\mathrm{i}}$] is constant during propagation in the IP medium. This is justified on the grounds that CMEs attain a constant mass relatively close to the Sun at $\approx$ $10~\mathrm{R_{\odot}}$ \citep{2010ApJ...722.1522V}. Therefore, the mass of the CME\,--\,sheath system $M_{\mathrm{tot}}$ varies with distance, and hence we have to generalize the 1D drag-force equation along the radial direction for a \textit{variable mass system}.

The force $F_{\mathrm{v m}}$ required to accelerate the fluid surrounding a moving submerged rigid body from \cite{book:1197664} is 
\begin{equation}
   F_{\mathrm{v m}} = \frac{\rho_{\mathrm{e}}\tau}{2}\left(\frac{\mathrm{D}V_{\mathrm{e}}}{\mathrm{D}t}-\frac{\mathrm{d}V_{\mathrm{i}}}{\mathrm{d}t}\right)
\end{equation}

with $\mathrm{\mathrm{D}/\mathrm{D}t}$ denoting the material derivative. Generalizing the virtual-mass force for a varying external (i.e. solar wind) density and an expanding CME structure, we get:

\begin{equation}
  F_{\mathrm{v m}} = \frac{\rho_{\mathrm{e}}\tau}{2}\left(\frac{\mathrm{D}V_{\mathrm{e}}}{\mathrm{D}t} \frac{\mathrm{d}V_{\mathrm{i}}}{\mathrm{d}t}\right) -u_{rel}\frac{\mathrm{d}}{\mathrm{d}t}\left(\frac{\rho_{\mathrm{e}}\tau}{2}\right),
\end{equation}

$u_{\mathrm{rel}}$ is the relative velocity of the body (CME) with respect to the fluid (solar wind). The origin of the notion of “virtual mass” becomes evident when we take a look at the momentum equation:

\begin{equation}
   \frac{\mathrm{d}P_{\mathrm{i}}}{\mathrm{d}t} = M_{\mathrm{i}}\frac{\mathrm{d}V_{\mathrm{i}}}{\mathrm{d}t} + u_{\mathrm{rel}}\cancel{\frac{\mathrm{d}M_{\mathrm{i}}}{\mathrm{d}t}} = F_{\mathrm{D}} + F_{\mathrm{vm}} \Rightarrow
\end{equation}

which becomes

\begin{equation}
   M_{\mathrm{i}}\frac{\mathrm{d}V_{\mathrm{i}}}{\mathrm{d}t}= F_{\mathrm{D}} + \frac{\rho_{\mathrm{e}}\tau}{2} \left(\frac{\mathrm{D}V_{\mathrm{e}}}{\mathrm{D}t}-\frac{\mathrm{d}V_{\mathrm{i}}}{\mathrm{d}t}\right) -u_{\mathrm{rel}}\frac{\mathrm{d}}{\mathrm{d}t}\left(\frac{\rho_{\mathrm{e}}\tau}{2}\right),
\end{equation}

with $P_{\mathrm{i}}$ corresponding to the momentum of a CME. Moving the derivative of the CME’s velocity from the right-hand side of the equation to the left and assuming a steady background solar wind, we get:

\begin{equation}
   \left(M_{\mathrm{i}}+\frac{\rho_{\mathrm{e}}\tau}{2}\right)\frac{\mathrm{d}V_{\mathrm{i}}}{\mathrm{d}t} = F_{\mathrm{D}} + \cancel{\frac{\rho_{\mathrm{e}}\tau}{2}\frac{\mathrm{D}V_{\mathrm{e}}}{\mathrm{D}t}} - u_{\mathrm{rel}}\frac{\mathrm{d}}{\mathrm{d}t}\left(\frac{\rho_{\mathrm{e}}\tau}{2}\right)
\end{equation}

The second term inside the parenthesis on the left-hand side of the above equation is the virtual mass. The virtual mass of a cylindrical CME is equal to $m_{virtual}=\frac{1}{2}\rho_{e}\pi R^{2}L$ (The added mass of a cylinder can be derived by considering a hydrodynamic force of an axisymmetric flow acting on it as it accelerates). The equation of motion then becomes:

\begin{equation}\label{eq:4}
   M_{\mathrm{tot}}\frac{\mathrm{d}V_{\mathrm{i}}}{\mathrm{d}t}= F_{\mathrm{D}} - (V_{\mathrm{LE}}-V_{\mathrm{e}})\frac{\mathrm{d}}{\mathrm{d}t}\left(\frac{\rho_{\mathrm{e}}\tau}{2}\right)
\end{equation}

In Equation \ref{eq:4} we introduced the speed of the CME leading edge [$V_{\mathrm{LE}}$] given mass pile-up occurs ahead of a CME which is faster than the ambient solar wind, as we study here. The second term of the right-hand side of Equation \ref{eq:4} is due to the relative motion of the CME with respect to the solar-wind flow. It is that $V_{\mathrm{LE}}=V_{\mathrm{i}}+V_{\mathrm{exp}}$, where $V_{\mathrm{exp}}$ is  the expansion speed of the cylindrical CME,

\begin{equation}\label{eq:5}
   V_{\mathrm{exp}}=\frac{\mathrm{d}R}{\mathrm{d}t}=\frac{\mathrm{d}R}{\mathrm{d}r}\frac{\mathrm{d}r}{\mathrm{d}t}=\frac{\mathrm{d}R}{\mathrm{d}r}V_{\mathrm{i}}
\end{equation}
Solving Equation \ref{eq:4} for $\mathrm{d}V_{\mathrm{i}}/\mathrm{d}t$ we obtain: 

\begin{equation}\label{eq:6}
   \frac{\mathrm{d}V_{\mathrm{i}}}{\mathrm{d}t}=\frac{F_{\mathrm{D}}}{M_{\mathrm{tot}}}-\frac{(V_{\mathrm{LE}}-V_{\mathrm{e}})}{M_{\mathrm{tot}}}\frac{\mathrm{d}m_{\mathrm{virtual}}}{\mathrm{d}t},
\end{equation}
and by using Equation \ref{eq:1} we finally obtain:
\begin{equation}\label{eq:7}
   \frac{\mathrm{d}V_{\mathrm{i}}}{\mathrm{d}t}=-\gamma C_{\mathrm{D}}|V_{\mathrm{i}}-V_{\mathrm{e}}|(V_{\mathrm{i}}-V_{\mathrm{e}})-\frac{(V_{\mathrm{i}}+V_{\mathrm{exp}}-V_{\mathrm{e}})}{M_{\mathrm{tot}}}\frac{\mathrm{d}m_{\mathrm{virtual}}}{\mathrm{d}t}
\end{equation}

Equation \ref{eq:7} describes the propagation of a faster than the ambient solar-wind CME subject to drag and incorporates mass pile-up. It is identical to the drag equations of \cite{2004SoPh..221..135C} and \cite{2010A&A...512A..43V}, with the addition of a second term in the right-hand side corresponding to a varying virtual mass. Note that the above studies neglected altogether the virtual mass. Both terms on the right-hand side of Equation \ref{eq:7}, are negative for fast CMEs, i.e. they cause a slow-down in the CME structure. The first term is associated with the aerodynamic drag and the second with the mass pile-up around the CME.

The numerical solution of Equation \ref{eq:7} requires the properties of the ambient solar wind. The solar-wind density is given by the \cite{1998SoPh..183..165L} empirical formula and its speed results from the application of the continuity equation $n_{\mathrm{e}}\, V_{\mathrm{e}}\,r^{2}=const$, describing the constant flow (conservation of mass flux) of solar-wind particles, where $n_{\mathrm{e}}$ is the number density of the solar wind and $r$ the heliocentric distance:

\begin{equation}\label{eq:8}
   n_{\mathrm{e}}V_{\mathrm{e}}r^{2}=const \Longrightarrow V_{\mathrm{e}}=\frac{n_{\mathrm{e}_{(1 \mathrm{AU})}}V_{\mathrm{e}_{(1 \mathrm{AU})}}(215~ \mathrm{R_{\odot}})^{2}}{n_{\mathrm{e}}r^{2}}
\end{equation}
For our study, the solar-wind electron-number density and speed at 1 AU were set as follows:
   $n_{\mathrm{e}_{(1 \mathrm{AU})}}= 7~\mathrm{cm^{-3}}$ and $V_{\mathrm{e}_{(1 \mathrm{AU})}}= 400~\mathrm{km\,s^{-1}}$, respectively. 

\subsection{Radial Profiles of Physical Parameters for Non-Eroded CMEs}

For non-eroded CMEs, it is reasonable to expect that their mass and magnetic flux should not vary with distance. This means that we need to consider in the numerical solution of Equation \ref{eq:7}, the radial profiles of CME density, radius, and (axial) magnetic field leading up to approximately constant CME mass and magnetic flux with distance.

There exist several works deducing the radial profiles of various CME physical parameters \citep[e.g.][]{1998AnGeo..16....1B,2005P&SS...53....3L, 2005JGRA..11010107W,2006SSRv..123..383F,2007JGRA..112.6113L}. They are mainly based on CME observations by the \textit{HELIOS} mission in $\approx$ 0.3 \,--\, 1 AU and describe the evolution of several CME physical parameters as power-laws of the radial distance. We then have that the CME parameter $y$ is given by $y=Ar^b$, with $A$ and $b$ corresponding to the constant and index of the power law, respectively. Calculating average values for the constants and the indexes of the power laws describing the CME radius and density of the mentioned-above works we have that:

\begin{equation}\label{eq:9}
    R=0.138\, r^{0.69}~\mathrm{AU} ~~~~ \mathrm{and} ~~~~  n_{\mathrm{i}}=6.59\,r^{-2.384}~\mathrm{cm^{-3}}
\end{equation}

which, under the assumption of a cylindrical CME, leads to an almost constant CME mass as a function of distance (i.e. $M_{\mathrm{i}}\propto r^{-0.004}$).

Following a similar procedure as above, this time for the CME magnetic-field magnitude, and the works of \citet{2005P&SS...53....3L}, \citet{2005JGRA..11010107W}, \citet{2006SSRv..123..383F} and \citet{2007JGRA..112.6113L}, we end up with the following power-law for $B_{i}$: 

\begin{equation}\label{eq:10}
   B_{\mathrm{i}}=11.4\,r^{-1.383}~\mathrm{nT}.
\end{equation}

Combining this power-law with the corresponding power-law for the CME radius (i.e. Equation \ref{eq:9}), we end up, for cylindrical CMEs, with a practically constant CME magnetic flux with distance, i.e. $\Phi_{\mathrm{B}}\propto r^{-0.003}$. In both Equations, \ref{eq:9} and \ref{eq:10}, the radial distance $r$ is expressed in AU.

\subsection{Aerodynamic Drag Force Acting on Eroded CMEs}

We now consider the impact of magnetic erosion on the drag force acting on CMEs. This is done by essentially assuming that erosion removes part of the mass and magnetic flux of the CME. 

Consider Figure \ref{fig:1} depicting the propagation of a cylindrical CME in the radial direction. If the CME encounters oppositely directed IMF, then reconnection occurs and progressively peels off concentric shells from the CME. This decreases both the mass and the magnetic flux of the structure. Given now that typically when magnetic reconnection occurs in simple geometries, we have that post-reconnection magnetic-field lines, and therefore the entrained mass as well, move perpendicularly to the inflow direction (e.g. check Panels b and c of Figure \ref{fig:1}), the addition of extra terms into Equation \ref{eq:7} associated with the momentum of the outflowing plasma is not warranted. As a matter of fact, in-situ observations of reconnection jets (``CME exhausts") occurring around or within CMEs showed that they mostly lie in planes that are perpendicular to the radial direction \citep{2007JGRA..112.8106G}.

The strength of this CME\,--\,IMF magnetic reconnection, and hence of the erosion that a CME undergoes, depends on the associated reconnection rate. The reconnection rate is proportional to the Alfvén speed, which is much higher near the Sun \citep[e.g.][]{2017SSRv..212.1159M}. Given now that the postulated reconnection involves two systems (i.e. CME and IMF) with different physical properties (e.g. densities, magnetic fields, etc.) estimating the magnetic-reconnection rate from the classical Sweet\,--\,Parker or Petschek reconnection models, based on symmetric inflow conditions, is not appropriate. We, therefore, opted to use a hybrid magnetic-reconnection rate derived by \cite{2007PhPl...14j2114C}.

\begin{equation}\label{eq:11}
   \Re=CS,
\end{equation}

where $C$ is a dimensionless coefficient that depends on the geometry of the magnetic reconnection process ($\approx 0.1$) and $S$ is a hybrid Alfvén speed multiplied by a hybrid magnetic-field strength  as deduced in \cite{2008JGRA..113.7210B}:

\begin{equation}\label{eq:12}
   S=2(B_{1}B_{2})^{3/2}(\mu_{0} \rho_{1} B_{2} + \mu_{0} \rho_{2} B_{1})^{-1/2}(B_{1}+B_{2})^{-1/2}.
\end{equation}

In the above equation, $B$ and $\rho$ are the magnetic field and the mass density respectively for the two different inflowing systems, with subscripts 1 and 2, corresponding to the CME and the ambient solar wind and IMF at the CME's front position, respectively. Equations \ref{eq:9} and \ref{eq:10}, the \cite{1998SoPh..183..165L} density profile and the azimuthal component of the IMF from \cite{1958ApJ...128..664P} supplied the CME and IMF\,--\,solar-wind parameters required for the calculation of $S$. Finally, $\mu_{0}$ is the magnetic permeability of vacuum.
 
Having established a means to calculate the CME\,--\,IMF magnetic reconnection rate as a function of distance in the inner heliosphere, we are now in a position to incorporate the impact of magnetic erosion on the CME kinematics. Magnetic erosion essentially boils down to a reduction of the CME radius at any given distance, compared to its value when no erosion occurs, and it is written as:

\begin{equation}\label{eq:13}
   R_{\mathrm{i}}=R_{\mathrm{i-1}}^{*}\,\left(\frac{\Re_{\mathrm{i}}}{\Re_{\mathrm{i-1}}}\right)^\alpha.
\end{equation} 

In essence, Equation \ref{eq:13} supplies the radius $R$ of the CME at a given radial position $\mathrm{i}$, by reducing its value [$R_\mathrm{{i-1}}^{*}$] at the previous radial grid position with index $\mathrm{i-1}$, calculated when no erosion occurs (i.e., following Equation \ref{eq:9}), by a factor depending on the magnetic reconnection rate $\Re$ at positions $\mathrm{i-1}$ and $\mathrm{i}$. The magnitude of the erosion, and therefore its impact on R, is controlled by the exponent $\alpha$. 

To determine $\alpha$, we considered a total CME magnetic-flux reduction from $20~\mathrm{R_{\odot}}$ (i.e. the starting distance of the application of the drag-force model discussed in the next Section) to 1 AU, of 20\,\%, 40\,\%, and 50\,\% consistent with CME observations showing evidence of magnetic erosion \citep{2015JGRA..120...43R, 2020GeoRL..4786372P}, leading to $\alpha$ values of 0.038, 0.089, and 0.121, respectively.

Magnetic erosion is then incorporated into drag-based CME kinematics (i.e. Equation \ref{eq:7}), via the associated reduction in CME radius (i.e. Equation \ref{eq:13}). Note, that the above effect, i.e. radius reduction, does not only have a geometrical impact but also leads to a decrease in the CME mass $\propto R^{2}$. Therefore, our treatment of magnetic erosion leads to a reduction in the structure's size along with an abatement of the virtual mass piling up in front of it. We will discuss in detail the implications of our prescription of the impact of erosion on CME radius in Section 4.

\section{Results}
Having presented in the previous section our theoretical framework dealing with both eroded and non-eroded CMEs, we will now apply it in order to study the structure's behavior. The propagation is studied via the numerical solution of Equation \ref{eq:7} from $x_{0}=20~\mathrm{R_{\odot}}$ to $215~\mathrm{R_{\odot}}$, i.e. 1 AU. We considered four cases: a CME without erosion, and three CMEs experiencing erosion leading to total magnetic-flux reduction between $x_{0}$ to 1 AU of 20\,\%, 40\,\%, and 50\,\% (see the previous section). Typically, applications of the drag model use the same $x_{0}$ as starting distance \citep[check for example Table 1 of][]{2021FrASS...8...58D}. The initial mass of the modeled CMEs was 1.74 $\times$ ${10}^{12}$ kg, consistent with observed distributions of CME masses in the corona \citep[e.g.][]{2010ApJ...722.1522V}, and their initial radius was $\approx 5.75~\mathrm{R_{\odot}}$, consistent with forward modeling of CMEs observed by STEREO \citep[e.g.][]{2009SoPh..256..111T}. In addition, all modeled CMEs have an angular width of $45^{\circ}$, consistent with STEREO observations of CMEs \citep{2009SoPh..256..111T}, which was kept constant during their propagation. This is valid even for eroded CMEs, given that as discussed in the previous section, in the frame of our model erosion affects only the radii of the postulated cylindrical CMEs and not their heights. Finally, the initial CME bulk speed was 1000 $\mathrm{km\,s^{-1}}$, therefore we were dealing with fast CMEs.

\subsection{Impact of Magnetic Erosion on CME--IMF Reconnection rate and CME Mass and Radius}

\begin{figure}[H]
    \centering
    \includegraphics[width=12cm]{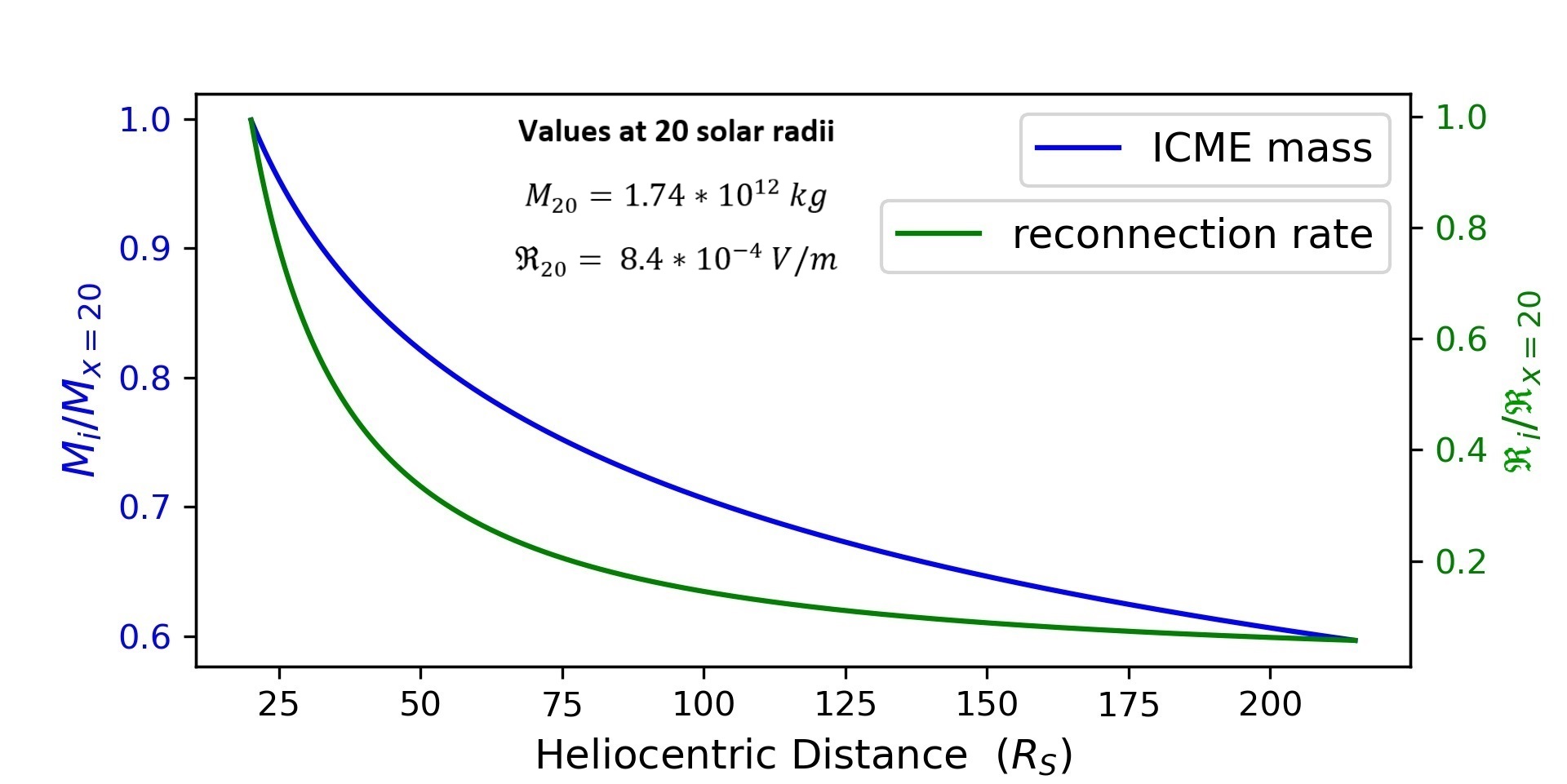}
    \caption{Radial evolution of the normalized CME\,--\,IMF magnetic reconnection rate (\textit{green line}) and the mass of the magnetic-flux rope (\textit{blue line}) for a CME undergoing a total magnetic-flux reduction of 40\,\% between $x_{0}$ and 1 AU. The i index (see the caption of the y-axis) captures the distance of the magnetic structure in units of solar radii from $x_0$ ($x=20$ $\mathrm{R_{\odot}}$) up to 1 AU ($x=215$ $\mathrm{R_{\odot}}$).}
    \label{fig:2}
\end{figure}

Before studying in detail the kinematics of eroded and non-eroded CMEs we give examples of the impact of erosion on pertinent CME characteristics. Figure \ref{fig:2} corresponds to a CME undergoing a total magnetic-flux reduction between $x_{0}$ and 1 AU of 40\,\%. It shows that the CME\,--\,IMF reconnection rate is stronger near the Sun and falls off rapidly with distance. This is expected on the grounds of higher Alfvén speeds close to the Sun. In addition, magnetic erosion leads to a significant decrease in the CME mass of the order 40\,\% from $x_{0}$ to 1 AU.

\begin{figure}[H]
    \centering
    \includegraphics[width=12cm]{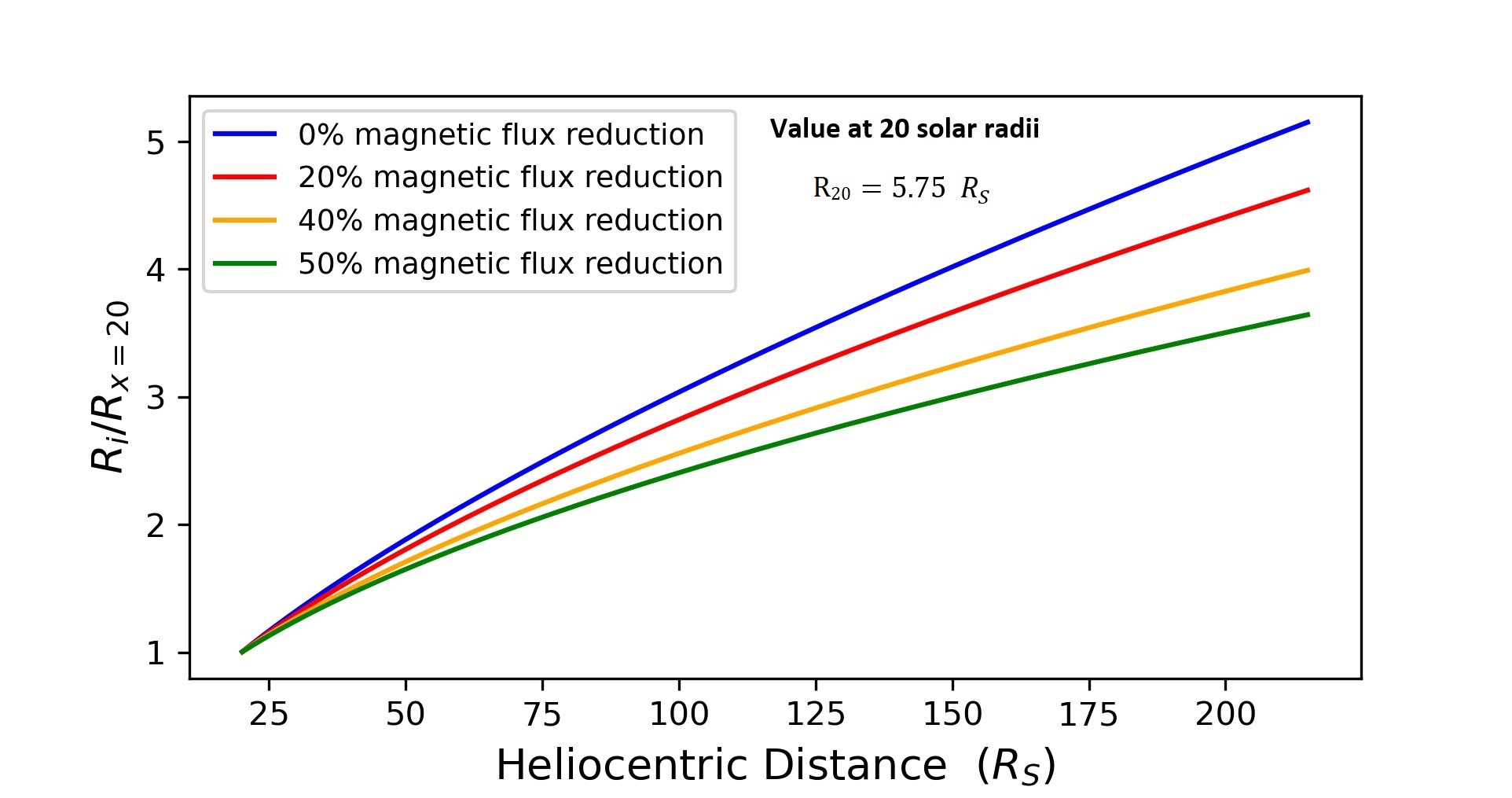}
    \caption{Radial evolution of the CME radius for a non-eroded CME (\textit{blue line}) and eroded CMEs with total magnetic-flux reduction of  20\,\%, 40\,\%, and 50\,\% (\textit{red, yellow}, \textit{and green lines}) between $x_{0}$ and 1 AU. The i index (see the caption of the y-axis) captures the distance of the magnetic structure in units of solar radii from $x_0$ ($x=20$ $\mathrm{R_{\odot}}$) up to 1 AU ($x=215$ $\mathrm{R_{\odot}}$).}
    \label{fig:3}
\end{figure}

Figure \ref{fig:3} contains the radial evolution of the CME radius for the non-eroded case, and three eroded cases with a 20\,\%, 40\,\% and 50\,\% magnetic-flux reduction. Once more the impact of erosion is obvious. The greater the magnetic-flux reduction, the smaller the increase of the CME radius. We have that the CME radius at 1 AU is smaller in comparison to its value for a non-eroded case, by factors of $\approx$ 3.5, 4.5, and more than 5 for magnetic-flux reduction of 20\,\%, 40\,\%, and 50\,\%, respectively.

\subsection{Kinematics of Eroded and Non-Eroded CMEs}

\begin{figure}[H]
    \centering
    \includegraphics[width=12cm]{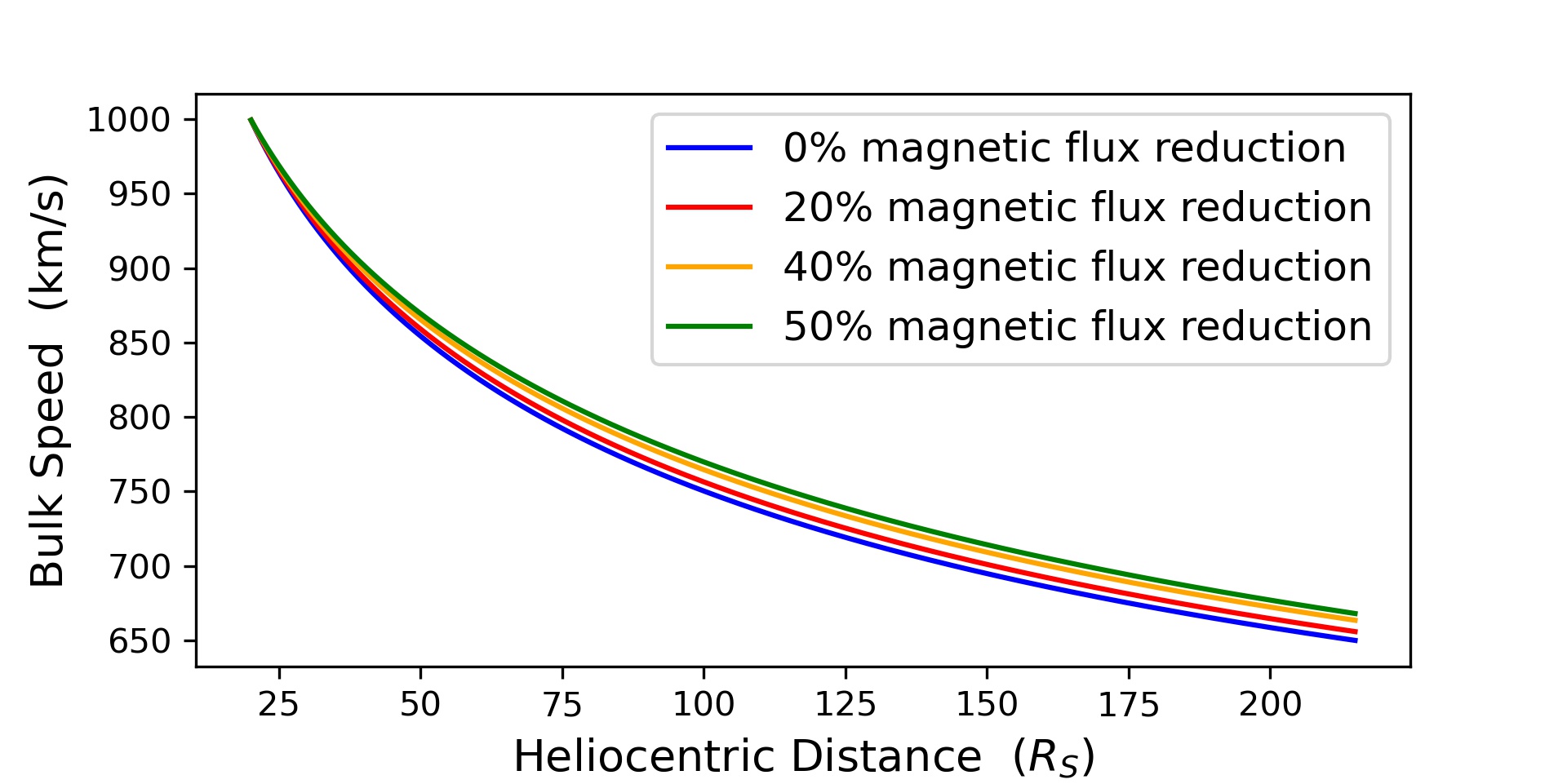}
    \caption{Radial evolution of CME bulk speed for a non-eroded CME (\textit{blue line}) and three eroded CMEs with total magnetic-flux reduction of  20\,\%, 40\,\%, and 50\,\% (\textit{red, yellow}, \textit{and green lines}) between $x_0$ ($x=20$ $\mathrm{R_{\odot}}$) and 1 AU ($x=215$ $\mathrm{R_{\odot}}$).}
    \label{fig:4}
\end{figure}

\begin{figure}[H]
    \centering
    \includegraphics[width=12cm]{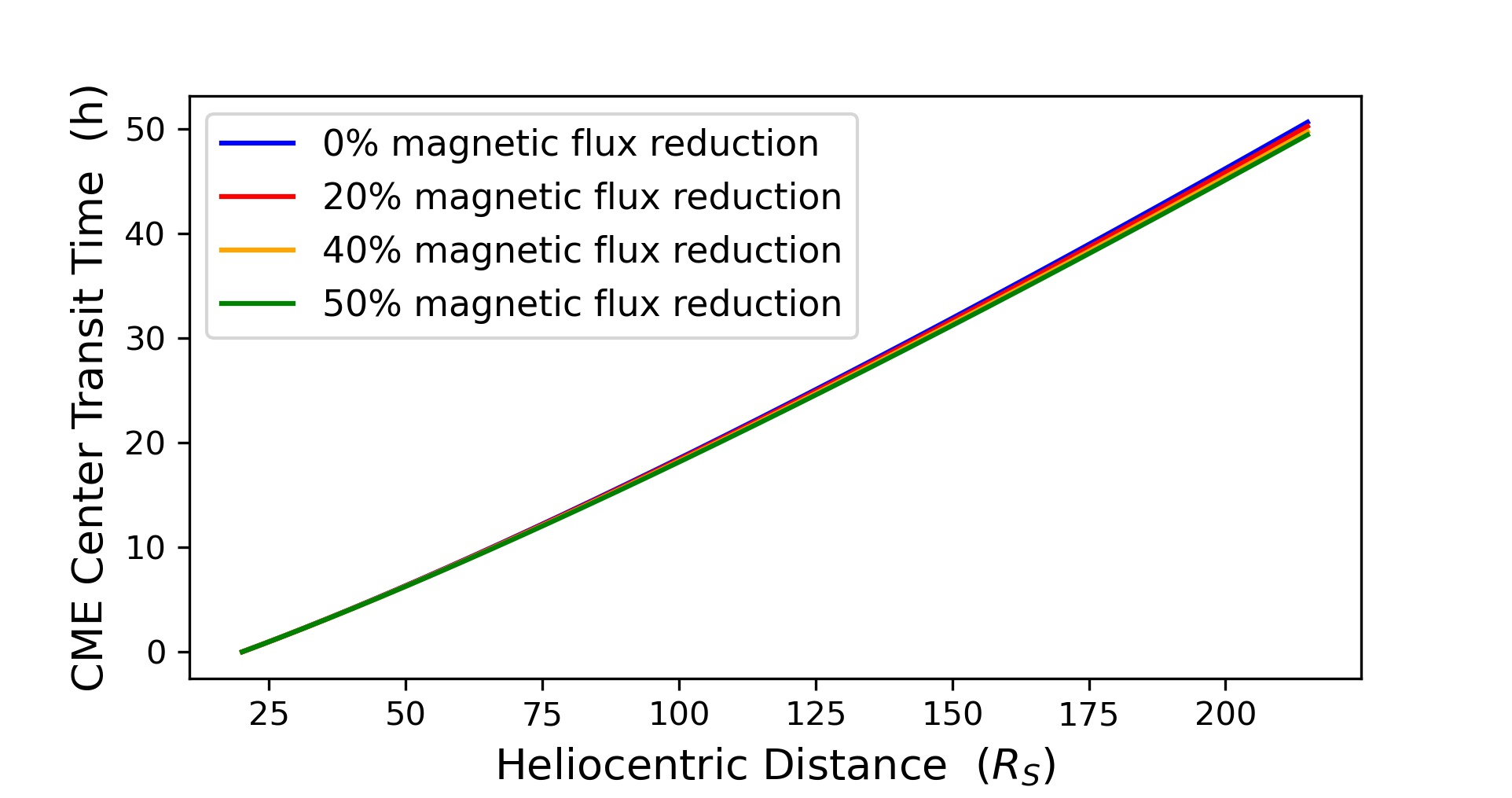}
    \caption{Transit time of CME center in the  $20-215~\mathrm{R_{\odot}}$ interval for a non-eroded CME (\textit{blue line}) and three eroded CMEs with total magnetic-flux reduction of  20\,\%, 40\,\%, and 50\,\% (\textit{red, yellow}, \textit{and green lines}) between $x_0$ ($x=20$ $\mathrm{R_{\odot}}$) and 1 AU ($x=215$ $\mathrm{R_{\odot}}$).}
    \label{fig:5}
\end{figure}

We now consider in detail the propagation of the four modeled CMEs. For this, we solved numerically the CME equation of motion (Equation \ref{eq:7}) along with Equation \ref{eq:5} describing the evolution of CME radius. Recall here that all four CMEs were identically initialized at $x_{0}$. Figure \ref{fig:4} and Figure \ref{fig:5} contain the CME bulk speed and transit time for the center of the structure, respectively, as a function of the heliocentric distance for the four modeled CMEs. The non-eroded case exhibits lower speeds than the eroded ones, and as a result, the CME center reaches 1 AU later, with the differences increasing with the magnitude of the erosion. The CME center reaches 1 AU at speeds of $\approx$ $6-18$ $\mathrm{km\,s^{-1}}$ higher and $0.38-1.21$ hours earlier with respect to the non-eroded case (Figure \ref{fig:5}).
 
\begin{figure}[H]
    \centering
    \includegraphics[width=12cm]{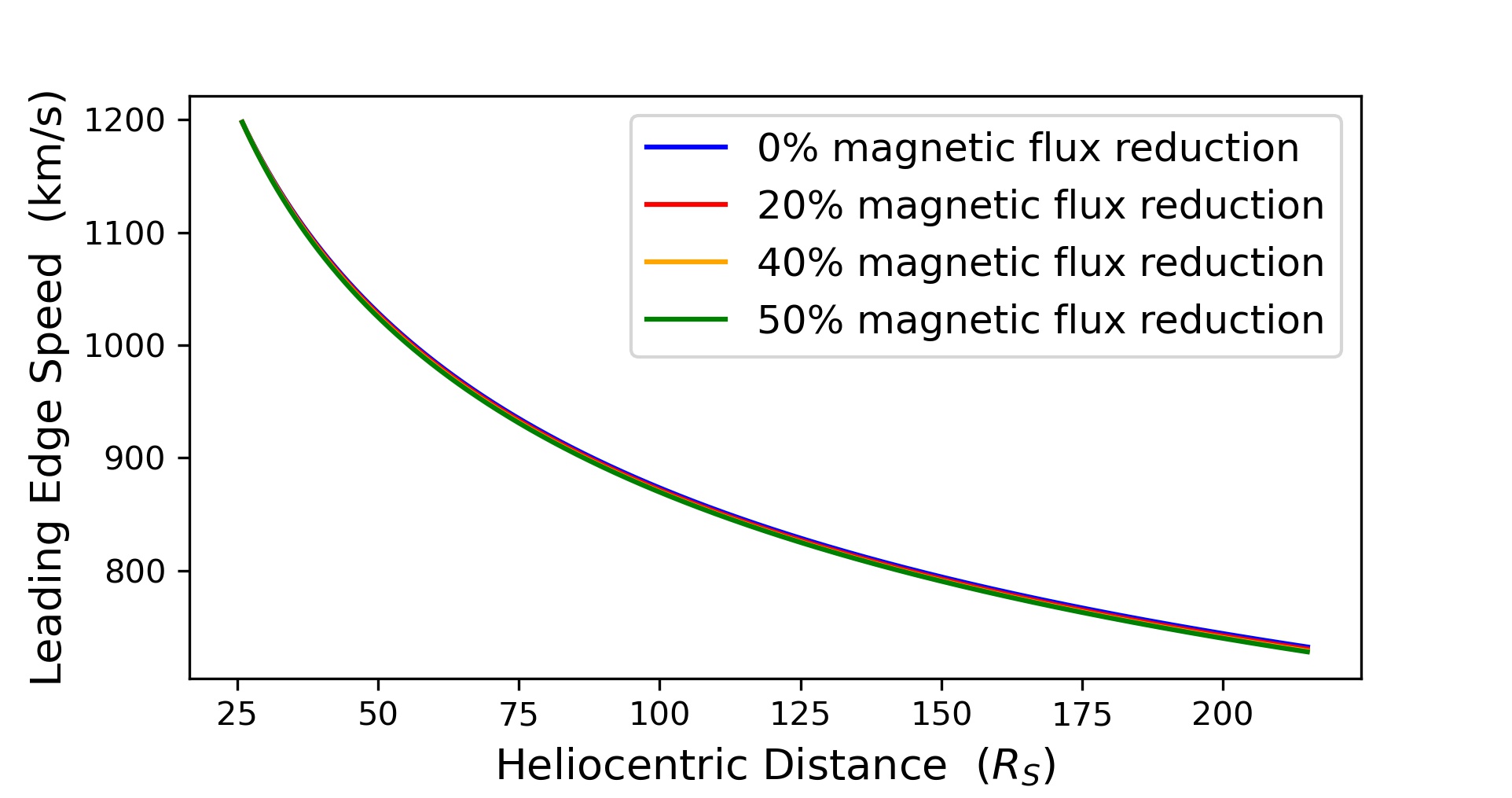}
    \caption{Radial evolution of CME leading-edge speed for a non-eroded CME (\textit{blue line}) and three eroded CMEs with total magnetic-flux reduction of  20\,\%, 40\,\%, and 50\,\% (\textit{red, yellow}, \textit{and green lines}) between $x_0$ ($x=20$ $\mathrm{R_{\odot}}$) and 1 AU ($x=215$ $\mathrm{R_{\odot}}$).}
    \label{fig:6}
\end{figure}

\begin{figure}[H]
    \centering
    \includegraphics[width=12cm]{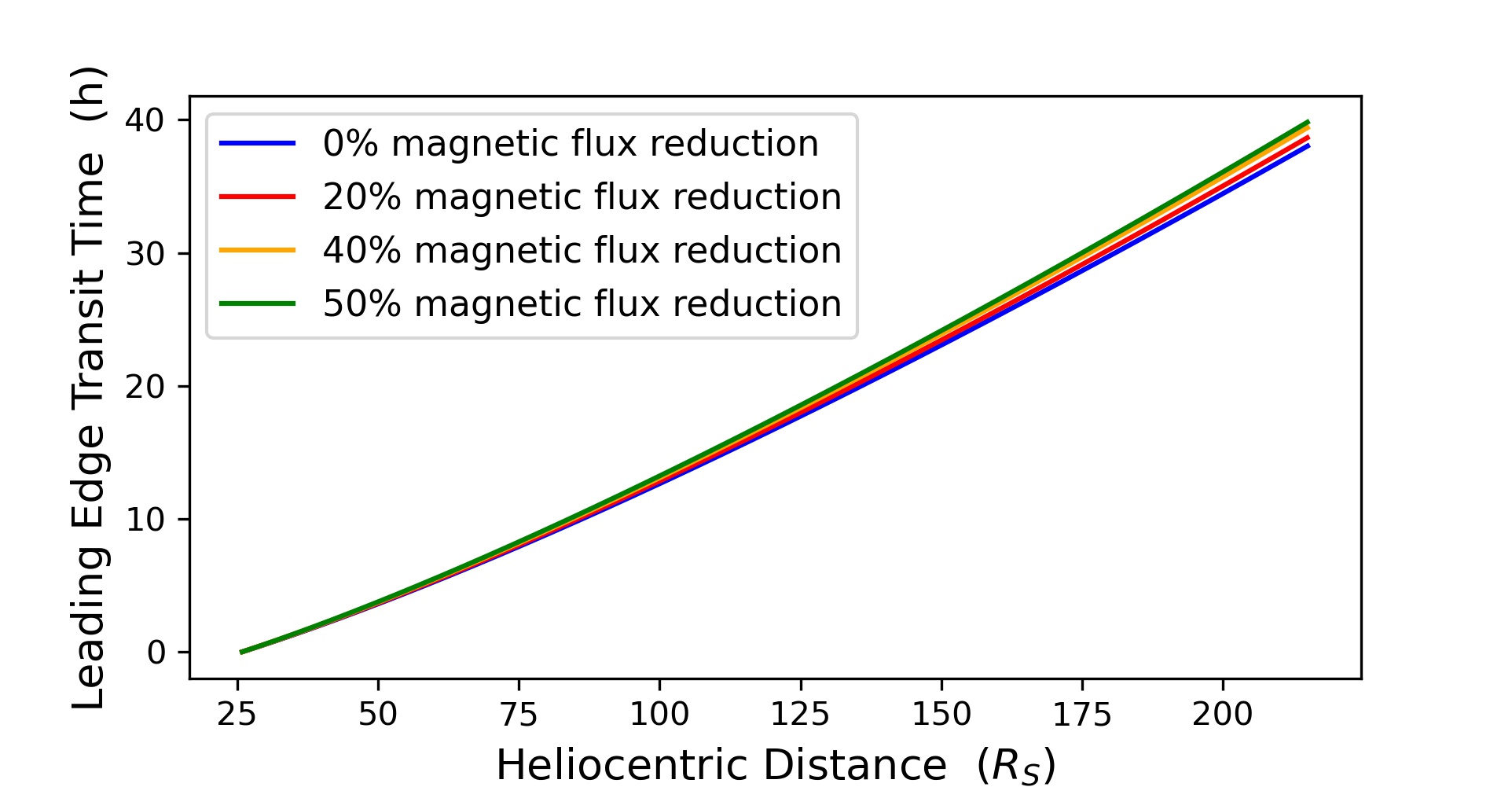}
    \caption{Radial evolution of CME leading-edge transit time for a non-eroded CME (\textit{blue line}) and three eroded CMEs with total magnetic-flux reduction of  20\,\%, 40\,\%, and 50\,\% (\textit{red, yellow}, \textit{and green lines}) between $x_0$ ($x=20$ $\mathrm{R_{\odot}}$) and 1 AU ($x=215$ $\mathrm{R_{\odot}}$).}
    \label{fig:7}
\end{figure}

By next adding the CME expansion speed to the CME bulk speed and the CME radius to the CME center radial distance, we obtain the CME leading-edge speed and heliospheric position respectively. Figures \ref{fig:6} and \ref{fig:7} contain the radial evolution of the CME leading-edge speed and transit time respectively. Depending on the magnitude of the erosion, the CME leading edge reaches 1 AU at speeds of $\approx$ $2-4.5$ $\mathrm{km\,s^{-1}}$ smaller and  $\approx$ $0.62-1.79$ hours later than the corresponding non-eroded case. 

From the above, and by juxtaposing Figures \ref{fig:4} and \ref{fig:5} with Figures \ref{fig:6} and \ref{fig:7}, we can readily reach an interesting conclusion, namely that magnetic erosion affects differently the CME center and leading edge, which are faster and slower respectively than these of corresponding non-eroded CME, respectively. To understand this behavior we have to investigate the varying impact of erosion on the CME bulk and expansion speeds.

\begin{figure}[H]
    \centering
    \includegraphics[width=12cm]{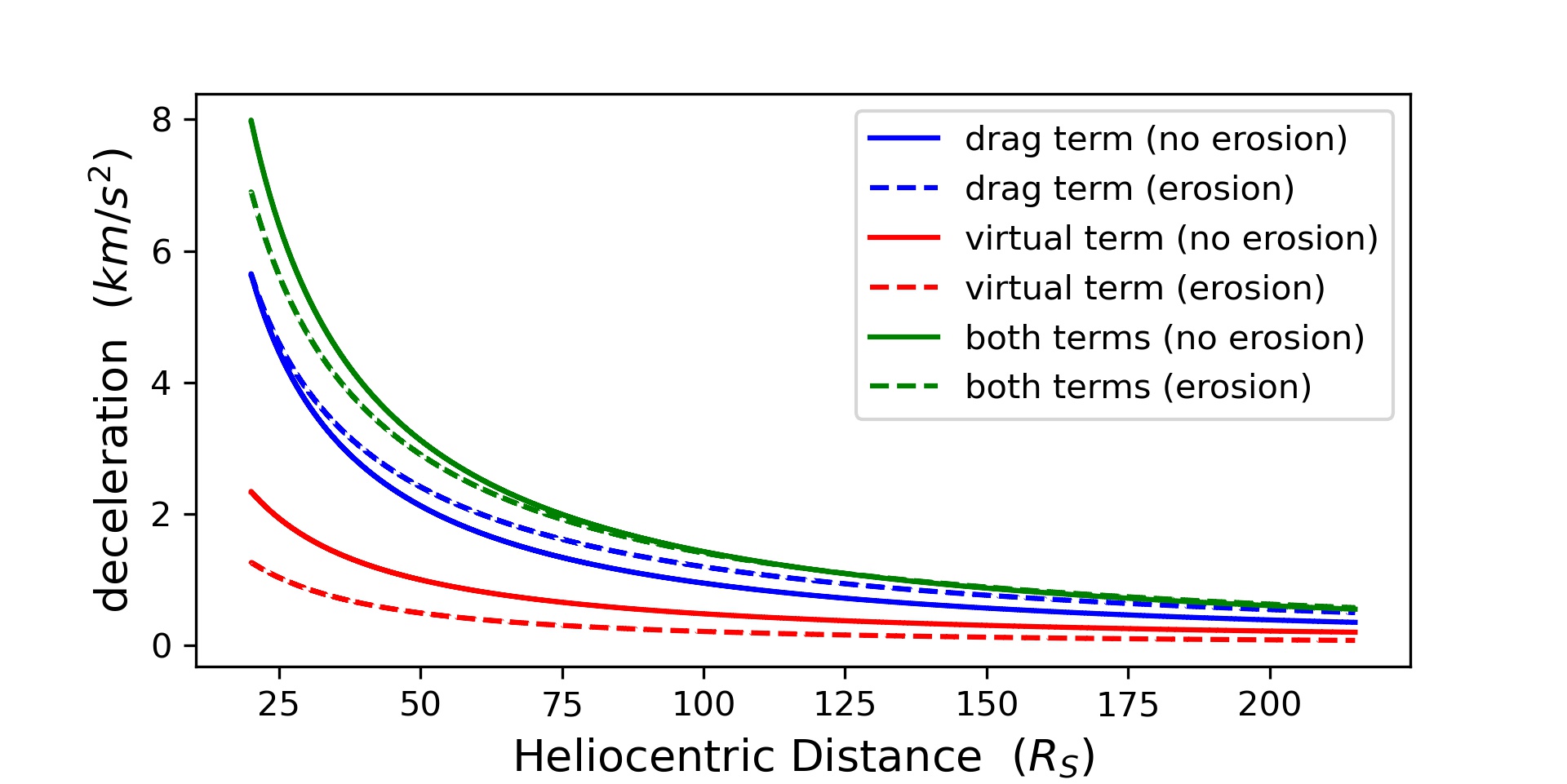}
    \caption{The deceleration terms (absolute values)
    of the CME bulk motion corresponding to the
    the drag force (first term of the right-hand side of Equation \ref{eq:7} - \textit{blue lines}) and the varying virtual mass (second term of the right-hand side of  Equation \ref{eq:7} - \textit{red lines}) for a non-eroded (\textit{solid lines}) and an eroded (\textit{dashed lines}) CME with 50\,\% magnetic-flux reduction. The \textit{green solid and dashed lines} correspond to the total deceleration
    (i.e. the sum of both terms on the right-hand side of 
    of Equation  \ref{eq:7}) acting on the eroded
    and corresponding non-eroded CME, respectively.}
    \label{fig:10}
\end{figure}

To better understand the impact of magnetic erosion on the CME bulk speed, we plot in Figure \ref{fig:10} the radial evolution of the two deceleration terms appearing on the right-hand side of Equation \ref{eq:7} for a non-eroded CME and the CME with maximum erosion, i.e. corresponding to a magnetic-flux reduction of $50\,\%$ from $x_0$ to 1 AU. We first note that the deceleration term due to the drag force (blue lines) is significantly higher than the varying virtual mass term (red lines). Their differences are more pronounced for distances of up to $\approx$ $100~\mathrm{R_{\odot}}$, and beyond this distance, both terms become progressively closer. Given that both terms are negative for the fast CMEs of our study, we conclude that the CME bulk speed is controlled in the same fashion by the drag and varying virtual mass. We next note that for the eroded CME, the deceleration due to the drag force is higher than that corresponding to the non-eroded CME while the opposite is true for the varying virtual mass. However, the sum of both deceleration terms (green lines) is smaller for the eroded CME of up to a distance of $\approx$ $100~\mathrm{R_{\odot}}$, and therefore the eroded CME experiences smaller deceleration compared to its non-eroded counterpart. This leads to a higher CME bulk speed for the eroded CME, and therefore to delayed CME-center transit time at 1 AU compared to that of the non-eroded CME.

On the other hand, since magnetic erosion essentially strips off concentric cells from CMEs (see, for example, the schematic of Figure 1), we could expect that an eroded CME would have a slower expansion speed compared to the corresponding non-eroded case (see also Figure \ref{fig:3}). Since now the CME leading-edge speed is essentially the sum of the CME bulk and expansion speed, which as seen above experience opposite dependencies on magnetic erosion, i.e. they increase (decrease) with respect to the non-eroded case, the leading-edge behavior is controlled by the competition between bulk and expansion speeds. For our studied cases, expansion speed is influenced stronger by erosion compared to the bulk speed and therefore eroded CMEs have delayed transit times of their leading edges compared to the corresponding non-eroded CMEs.

\subsection{Varying Initial CME Bulk Speed and Starting Distance of Erosion}
In this section, we perform two parametric studies of eroded and non-eroded CMEs by varying two of the parameters used in the initialization of the modeled CMEs. They pertain to the initial CME bulk speed and the starting distance of the application of the erosion to CMEs. 

\begin{figure}[H]
    \centering
    \includegraphics[width=12cm]{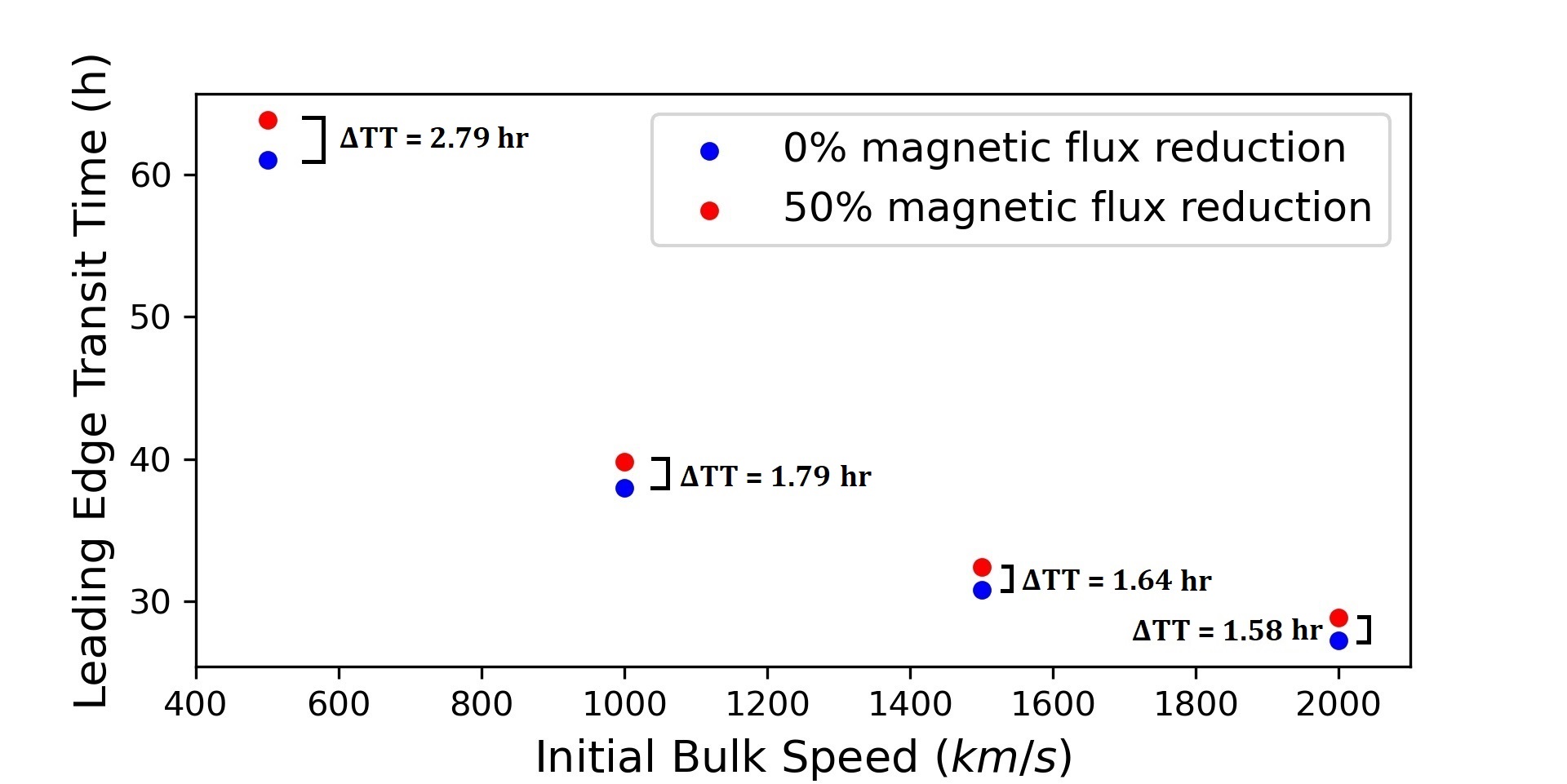}
    \caption{CME leading-edge transit time at 1 AU as a function of the initial bulk speed  at $\mathrm{R_{\odot}}=20$ for two different values of magnetic-flux reduction, 0\,\% (\textit{blue dots}) and 50\,\% (\textit{red dots}). The transit time difference between the eroded and the non-eroded case ($\Delta TT = TT_{\mathrm{eroded}} - TT_{\mathrm{non-eroded}}$) is given next to each \textit{pair of dots}}.
    \label{fig:8}
\end{figure}

Figure \ref{fig:8} contains the CME leading-edge transit time to 1 AU as a function of the initial bulk speed. We readily note that the higher the initial bulk speed, the lesser the impact of erosion, i.e. the smaller the difference in the transit time at 1 AU between an eroded and a non-eroded case. For the considered initial CME bulk speeds in the range of 500\,--\,2000 $\mathrm{km\,s^{-1}}$, the transit time difference between an eroded and non-eroded CME is $\approx$ 1.6\,--\,2.8 hours.

The decrease in the transit-time differences at 1 AU between an eroded and non-eroded CME with increasing initial bulk speed could be rather readily understood because we assumed drag-dominated CMEs. Magnetic erosion, when considered, is also incorporated into the drag prescription. Since drag essentially attempts to bring the structure's speed closer to the ambient solar-wind speed, a high initial bulk speed case is expected to be less affected by interactions with its environment compared to a lower speed case, and hence whether erosion occurs or not, has less impact on its kinematics.

\begin{figure}[H]
    \centering
    \includegraphics[width=12cm]{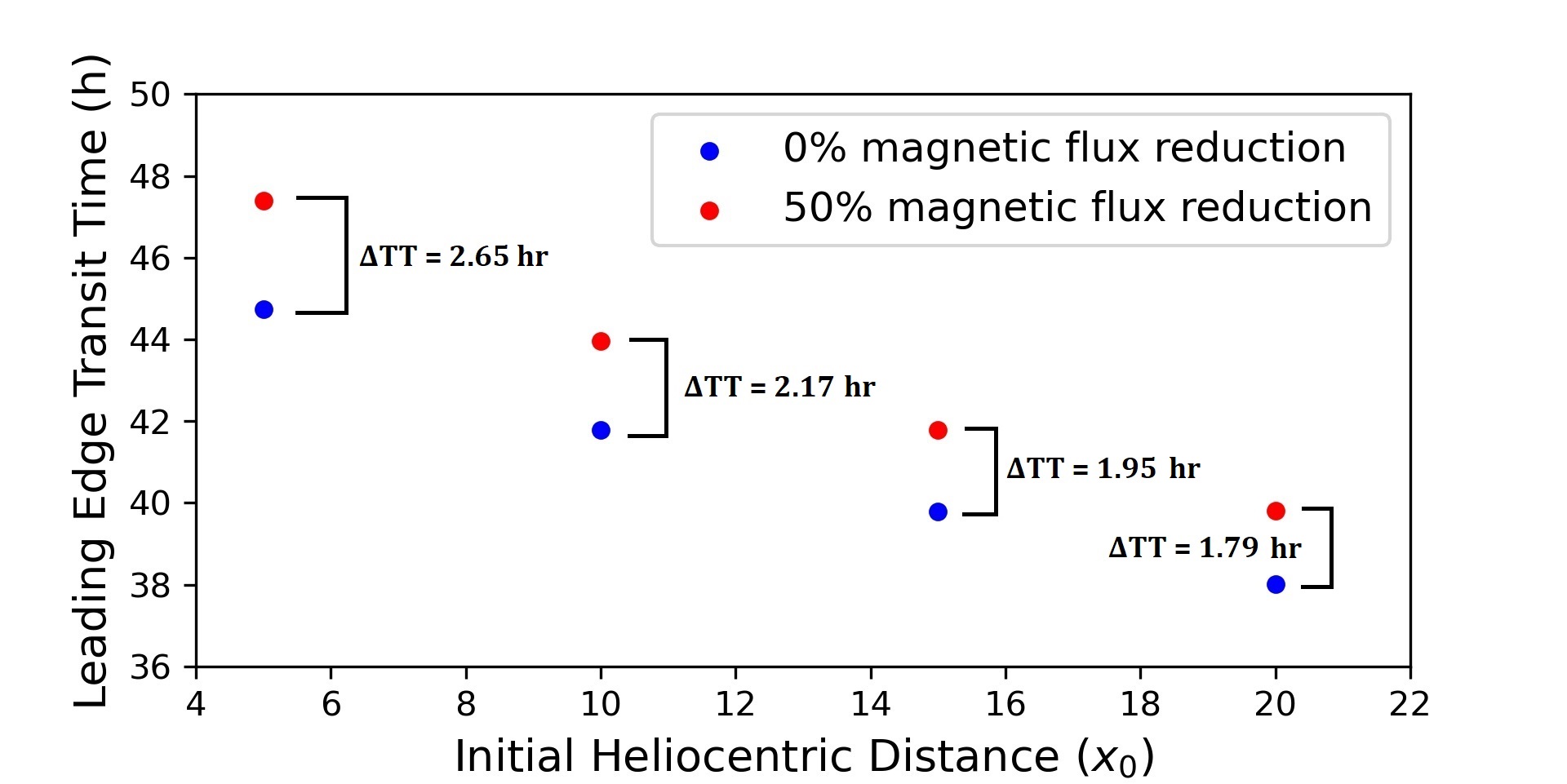}
    \caption{The transit time of a CME at 1 AU as a function of the starting distance of application of magnetic erosion and drag-based kinematics for two different values of magnetic-flux reduction, 0\,\% (\textit{blue dots}) and 50\,\% (\textit{red dots}). The transit-time difference between the eroded and the non-eroded case ($\Delta TT = TT_{\mathrm{eroded}} - TT_{\mathrm{non-eroded}}$) is given next to each pair of \textit{dots}}.
    \label{fig:9}
\end{figure}

Given now that the starting distance of the CME\,--\,IMF reconnection (i.e. the starting distance of magnetic-erosion application) for an eroded CME is rather unknown, and for simplicity, we assumed that it coincided with the starting distance $x_0$ of the CME drag-dominated propagation. We next investigated the impact of the (common) starting distance for magnetic erosion and drag-based CME dynamics on CME transit time at 1 AU. Figure \ref{fig:9}, illustrates the transit time at 1 AU as a function of $x_0$ for an eroded and a non-eroded event. For each starting distance, we calculated new $\alpha$ values (Equation \ref{eq:13}) so that the magnetic-flux reduction was 50\,\%, i.e. the same as in the considered case with maximum erosion discussed in the previous sections. From  Figure \ref{fig:9} we have that the transit time difference between an eroded and a non-eroded CME decreases from $\approx$ 2.6 down to 1.8 hours with $x_0$ taking values from the interval $5-20~\mathrm{R_{\odot}}$.

Overall, the decrease in the transit-time differences between eroded and non-eroded CMEs with increasing $x_0$ could be attributed to the fact that the reconnection rate sharply increases when approaching the Sun (e.g. Figure \ref{fig:2}). While the magnetic-flux reduction at 1 AU is the same for all studied cases, the non-identical reconnection rates applied to CMEs with different $x_{0}$ might be accountable for these differences. In other words, an eroded CME with a smaller $x_0$  would experience high reconnection rates close to the Sun, and hence the impact of erosion on its kinematics is expected to be more pronounced than for a CME with a larger $x_0$ which ``encounters" lower reconnection rates.

\section{Discussion and Conclusions}

In this work we developed a new drag-based model for the propagation of fast CMEs in the inner heliosphere that incorporates two significant additions over its predecessors, namely it includes the virtual mass via a variable-mass system formulation and CME magnetic erosion due to CME\,--\,IMF reconnection. In our model, magnetic flux and mass reduction due to magnetic erosion are controlled by the reconnection rate between the CME and IMF magnetic fields, which removes outer-shell mass from the cylindrical CME perpendicular to the propagation direction. Removing outer shells from the CME gives rise to a reduction of its magnetic flux as well, an essential attribute of eroded CMEs. Magnetic erosion influences the bulk and expansion speeds of the postulated cylindrical CME differently. While magnetic erosion speeds up the CME bulk speed with respect to a non-eroded case, its expansion speed slows down at a higher rate. As a net result, the eroded CME leading edge reaches 1 AU \textit{later} than its non-eroded equivalent. This delay depends on the strength of the erosion. 

Our results suggest that the addition of magnetic erosion into drag-based models has a significant impact on ToA predictions. Note here that our model is treating only the kinematics of the magnetic ejecta associated with CMEs and not their corresponding shocks and sheaths. Comparisons with actual in-situ CME ToA observations need to focus, therefore, on the ejecta (i.e. the magnetic obstacle). For the small number of cases we studied, we found that in the presence of erosion, a CME could reach 1 AU by up to $\approx$ 3 hours delay, with respect to the corresponding non-eroded case. Such delays represent a substantial fraction of the error in our existing time-arrival predictions of 10\,--\,12 hours as discussed for example in \citet{2019RSPTA.37780096V}. However, the majority of these predictions refer to shock/sheath ToAs rather than ToAs pertinent to the magnetic ejecta.

Drag-based models typically predict \textit{earlier} CME arrivals at 1 AU. For instance, \citet{2018ApJ...854..180D} discovered, for a large sample of CMEs analyzed, that the drag model leads to a mean ToA of -9.7 hours, with the minus sign corresponding to earlier predicted CME arrival at 1 AU as compared to the associated in-situ observations. Their earlier-than-observed predicted CME ToAs were particularly pronounced for fast CMEs \citep[check Figure 6a in][]{2018ApJ...854..180D}. Although the observed in-situ arrival times of the \citet{2018ApJ...854..180D} study were relevant to the CME shock and not the magnetic ejecta per se, our findings of delayed CME arrivals of eroded cases nevertheless suggest that adding erosion to the prescription of CME propagation could improve the drag-based predictions of fast CME arrivals at 1 AU.

Note here that the plasma and magnetic-field conditions
upstream of the modeled CMEs of our study, as required for the calculation of the CME\,--\,IMF reconnection rate, correspond to a quiescent inner heliosphere, i.e. without any large-scale transients as CMEs. However, CMEs compress and distort the sheath region which is the actual CME\,--\,``background" interface. Therefore, updated calculations of reconnection rates in CMEs incorporating the properties of CME sheaths would add additional realism to our model.

Recently \cite{2019A&A...632A..89H} performed 2.5D (axisymmetric) MHD simulations of CMEs and investigated the role of the polarity of the internal magnetic field in their evolution. They noticed that for the same initial conditions (e.g. CME speed and density, solar wind density, etc.) inverse CMEs (i.e. with the same magnetic-field polarity as in the IMF in front of them) reach 1 AU faster than normal CMEs (i.e. with opposite magnetic-field polarity compared to the IMF in front them). Therefore, when it comes to magnetic reconnection at the front of the magnetic structure, inverse (normal) polarity CMEs correspond to non-eroded (eroded) CMEs. The CMEs in their simulations were magnetized and dense plasma blobs.

For fast CMEs, with speeds in the inner corona ($2~\mathrm{R_{\odot}}$) of 800 and 1200 $\mathrm{km\,s^{-1}}$, \citet{2019A&A...632A..89H} concluded that the magnetic ejecta of eroded CMEs could be delayed from $\approx$ 0.5\,--\,1.5 hours with respect to a non-eroded CME, which is consistent with our results. The follow-up study of \citet{2021Geosc..11..314H} found that for CMEs undergoing erosion, both their mass and magnetic-flux decrease with distance, again along the lines of our model. The delayed ToAs of the magnetic ejecta of eroded CMEs of the \citet{2019A&A...632A..89H} simulations were attributed to magnetic reconnection occurring at the front of the CME which strips off magnetic shells from it, which leads in the CME frame, to a recession of its front. This is similar to the erosion-related decrease in the CME radius expansion of our work  on Equation \ref{eq:13}. The apparent similarities between our much-simplified model and the fully fledged MHD simulations of \citet{2019A&A...632A..89H} are encouraging and prompt for further analysis.

Based on our core hypothesis that magnetic erosion peels off the outer layers of CMEs, it is reasonable to expect that it could diminish the tension force exerted by the azimuthal magnetic field on CMEs. This reduction in tension could, in turn, lead to a decrease in the confinement of the CME internal plasma and magnetic field, resulting in an over-expansion (compared to a non-eroded case) of its cross-sectional area, which could affect its kinematics and ToA at 1 AU. This phenomenon is most likely to occur in regions closer to the Sun, where magnetic reconnection is facilitated by the higher Alfvén speeds. While erosion could potentially introduce an even greater imbalance between internal pressure and magnetic tension and consequently affect the transit times of CMEs at 1 AU, the MHD simulations presented in \citet{2019A&A...632A..89H, 2021Geosc..11..314H} suggest that the postulated erosion-related CME over-expansion does not seem to have a major impact on the kinematics of eroded CMEs, which as discussed in the previous paragraph reach 1 AU later than their non-eroded counterparts.

Moreover,
\citet{2009A&A...498..551D} studied the causes of CME expansion with analytical solutions. They modeled the evolution of cylindrical flux ropes as a series of force-free field states with ideal MHD and minimization of the magnetic energy with conserved magnetic helicity with the latter reproducing a situation for which the CME undergoes reconnection. Although they concluded that the ambient pressure has the most significant impact on the expansion, the various cases slightly deviated from each other due to the different magnetic field configurations and evolution, with the reconnection case exhibiting a shallower radius expansion rate in the inner heliosphere (check Figure 4 in \citet{2009A&A...498..551D}), in qualitative agreement with our results.

Magnetic erosion could also influence CME kinematics by modifying the exerted Lorentz force, particularly closer to the Sun. The effect of erosion on the acceleration of a CME is not straightforward, and it depends on various factors. However, stripping away the outer azimuthal magnetic field of a CME due to erosion could impact its early kinematics. The magnetic field plays a central role in driving the acceleration of a CME through the Lorentz force. Erosion-induced weakening of the magnetic field of CMEs could potentially alter their acceleration, with a more pronounced effect expected in regions closer to the Sun. However, investigating the impact of erosion on the Lorentz force in the proximity of the Sun is outside the scope of this study, which focuses on the drag-based kinematics of CMEs. This is because drag forces increasingly dominate the kinematics of fast CMEs beyond 15 solar radii, as reported by \citet{ 2015ApJ...809..158S}. Thus, further research is necessary to determine the precise effect of erosion on the acceleration of CMEs closer to the Sun by exploring the impact of magnetic erosion on the Lorentz force.

Although the core of this work was the incorporation of magnetic erosion into drag-based CME kinematics, our results have broader implications for the impact of magnetic erosion on CMEs. In Section 2.3, we laid down a general reconnection-based framework to figure out how magnetic erosion influences the CME radius, and accordingly the CME mass, by considering Equations \ref{eq:9} and \ref{eq:11}\,--\,\ref{eq:13}. Therefore, our prescription of CME mass evolution under the influence of erosion is not directly linked to the drag-based CME kinematics (or any other CME propagation model), and hence it represents a generic prediction of the effect of magnetic erosion on CME mass depending only on the specifics of the employed reconnection model. The significant decrease of the CME mass of the eroded CME of Figure \ref{fig:2} from $25~\mathrm{R_{\odot}}$ to 1 AU, is stronger within $\approx$ $25-75~\mathrm{R_{\odot}}$. This interval is covered almost exclusively by the heliospheric imagers of STEREO which however lack the required sensitivity to fully distinguish CMEs from their sheaths. This prohibits detailed CME mass observations to be compared against our predictions. Placing the starting distance of magnetic erosion application, i.e. $x_{o}$, closer to the Sun shifts the most noteworthy erosion-related CME mass depletion deeper within the corona, namely in the field-of-view (FoV) of the LASCO C2 and C3 coronagraphs where CME mass observations abound. LASCO observations of the mass evolution of more than 10,000 events found that, on average, CMEs reach a constant mass above around $10~\mathrm{R_{\odot}}$ \citep{2010ApJ...722.1522V}. Note here that these observations are not directly comparable with our model predictions since \citet{2010ApJ...722.1522V} considers both the upper part and legs, while our study considers only the CME upper part. On the other hand, measurements of the mass evolution of the upper parts (fronts) of a small sample of 13 CMEs within the LASCO FoV found no evidence of pileup, contrary to the general expectation  \citep{2018SoPh..293...55H}. The authors attributed the lack of pileup to the sensitivity of the observations but our modeling here suggests that erosion could be a factor. It is therefore unclear whether mass measurements are at odds with magnetic erosion initiated within the LASCO-C2 and LASCO-C3 FoV.

The focus of our study was on fast CMEs, i.e. CMEs with speeds above the speed of the ambient solar wind. For slow CMEs, an additional complication arises from the fact, that both their front and rear parts could exhibit mass pile-up. For instance, and for particular values of the bulk and expansion speed of a slow CME, it is possible that its front (rear) part could be faster (slower) than the ambient solar wind, and therefore, pile-up could occur on both parts. On the other hand, for fast CMEs, both front and rear parts are expected to be faster than the ambient solar wind; therefore, pile-up occurs only in the front. Therefore, more detailed studies of the mass/density evolution of CMEs fronts are needed to understand the impact of magnetic erosion on the kinematics of slow CMEs. In addition, our model may be extended to non-cylindrical geometries (e.g. spheromak, 3D flux rope).

Deducing power laws for the physical properties of eroded and non-eroded CMEs represents a crucial task for future improvement of our model, where data from the \textit{Parker Solar Probe} and \textit{Solar Orbiter} missions that cover both remote sensing and in-situ the corona and the inner heliosphere could be utilized. In addition, we used power-laws of CME properties based on \textit{HELIOS} observations, which are valid from 0.3 AU outwards. Therefore, it is essential to extend such power laws ``deeper" in the corona with \textit{Parker Solar Probe} observations. 

The most obvious extension of our model is to apply it to actual CME events undergoing erosion. Given that our study did not aim to model specific CMEs, we assumed various starting distances for the application of drag and erosion to the CME. However, for real events, the starting distance could be derived using the methodology of \citet{2015ApJ...809..158S}, which allows the determination of the distance at which the drag force dominates over the Lorentz force. Finally, given that our model depends on several, mainly empirically deduced parameters, concerning the properties of the studied CME including the amount of magnetic-flux erosion, as well as the properties of the background solar wind and IMF, the introduction of probability distributions for its input parameters instead of single values as input will allow for an estimation of the ToA forecast uncertainty \citep[e.g.][]{2018JSWSC...8A..11N}.

As aforementioned, we need to conduct additional comparisons between our  model and 3D MHD simulations to shed more light on the complicated processes (i.e. magnetic erosion and pile-up) that underpin the behavior of our model.

\section{Acknowledgments and Funding}
The authors acknowledge the reviewer for constructive comments and suggestions. S. Stamkos was supported by the project “Dioni: Computing Infrastructure for Big-Data Processing and Analysis.” (MIS No. 5047222) which is implemented under the Action “Reinforcement of the Research and Innovation Infrastructure”, funded by the Operational Programme ``Competitiveness, Entrepreneurship and Innovation" (NSRF 2014-2020) and co-financed by Greece and the European Union (European Regional Development Fund). A. Vourlidas was supported by NASA grant 80NSSC22K0970. S. Patsourakos acknowledges support by the ERC Synergy Grant (GAN: 810218) ‘The Whole Sun’. We thank G. Napoletano for useful discussions.

\bibliographystyle{spr-mp-sola}
\bibliography{sola_bibliography}

\end{article} 

\end{document}